\documentclass[
  aps,
  prd,
  11pt,
  superscriptaddress,
  notitlepage,
  onecolumn,
  nofootinbib
]{revtex4-1}

\AtBeginDocument{
  \linespread{1.05}\selectfont
}

\usepackage[
  a4paper,
  left=2.3cm,
  right=2.3cm,
  top=2.5cm,
  bottom=2.5cm
]{geometry}

\usepackage{microtype}
\usepackage{graphicx}
\usepackage{orcidlink}
\usepackage{enumerate}
\usepackage{enumitem}
\usepackage{ulem}
\usepackage{amssymb}
\usepackage{tikz}
\usepackage{mathtools}
\usepackage[dvipsnames]{xcolor}
\usepackage{ytableau}
\usepackage{soul}
\usetikzlibrary{arrows.meta,positioning,fit,calc}

\definecolor{linkcolor}{rgb}{0.0,0.3,0.5}
\definecolor{mypurple}{RGB}{143,116,210}

\graphicspath{{Figs/}}

\newcommand{\hias}{
School of Fundamental Physics and Mathematical Sciences,
Hangzhou Institute for Advanced Study,
University of Chinese Academy of Sciences,
Hangzhou 310024, China
}

\begin{document}


\title{Relative-Phase Control of Common-Bridge Survival in Collisions of Radially Excited Boson Stars}

\author{Bo-Xuan Ge
\orcidlink{0000-0003-0738-3473}}
\email{bo-xuan.ge@ucas.ac.cn}
\affiliation{\hias}


\title{Relative-phase control of common-bridge survival in collisions of radially excited boson stars}

\begin{abstract}
We show that the relative phase primarily controls the survival, rather
than the initial formation, of the common bridge in head-on collisions
of radially excited boson stars. In the benchmark configuration, all
sampled phases form both a common and a strong bridge. After aligning the
evolutions at the same operational weak-contact level, their early
post-contact evolution remains similar, while a clear phase hierarchy
develops only later, with larger phase offsets producing a weaker and
less persistent connection. The same suppression persists under
controlled variations of the central scalar amplitude, quartic
self-interaction strength, and scalar potential. These results identify
relative phase as a control on the nonlinear persistence of the common
bridge rather than simply on the instantaneous scalar-field overlap.
\end{abstract}

\maketitle

\section{Introduction}
\label{sec:introduction}

Bosonic fields motivated by particle physics and cosmology can form
self-gravitating compact configurations. Boson stars, composed of
complex scalar fields, are among the simplest examples and have been
considered in connection with dark matter and as alternatives to black
holes~\cite{Liebling:2012fv,Visinelli:2021uve,Widdicombe:2018oeo}.
They were first constructed by Kaup and by Ruffini and
Bonazzola~\cite{Kaup:1968zz,Ruffini:1969qy}, and self-interactions can
substantially increase their characteristic mass
scale~\cite{Colpi:1986ye}. Their equilibrium structure, stability, and
dynamics have since been studied in a broad range of models; see
Refs.~\cite{Schunck:2003kk,Liebling:2012fv} for reviews. Bosonic compact
objects can also form dynamically through scalar-field relaxation and
gravitational cooling~\cite{Seidel:1993zk}, with related formation and
collapse scenarios explored for axion stars and
oscillatons~\cite{Helfer:2016ljl,Widdicombe:2018oeo,Widdicombe:2019woy}.

Radially excited boson stars provide a useful setting in which to study
how field degrees of freedom interact with internal scalar structure.
Their profiles contain radial nodes, and their stability differs from
that of fundamental configurations. Early nonlinear evolutions found
excited states to be dynamically fragile~\cite{Balakrishna:1998Excited}.
Self-interactions can extend their lifetime in spherical
dynamics~\cite{SanchisGual:2021Excited,Brito:2023Excited}, while
multidimensional evolutions reveal additional instability channels and
nonlinear morphologies~\cite{Brito:2026Chains}. Recent work has further
examined their radial spectra and dynamical
signatures~\cite{Hao:2026ogd}. Collisions of excited configurations can
produce transient chain- and ring-like
structures~\cite{Brito:2026Chains,Ge:2026wzh}.

It is useful to distinguish two features of these evolutions from the
outset. A common bridge refers to a finite-amplitude connection between
the two objects, whose connectivity can be quantified independently of
the detailed morphology. Its definition does not rely on radial
excitation. A chain, by contrast, refers here to the multi-peak spatial
structure that can develop in collisions of radially excited boson
stars. We therefore treat these two aspects separately. The
density-based bridge diagnostic \(B\) measures the connectivity of the
central region, while \(S_{\rm chain}\) characterises the instantaneous
chain-like morphology. The latter does not define either the formation
or the survival of the common bridge.

Binary boson stars possess an additional degree of freedom that is
absent from black-hole binaries and ordinary fluid-star binaries: the
relative phase of the underlying complex scalar fields. Early head-on
simulations showed that otherwise identical boson stars with different
phases can follow different merger dynamics and radiate
differently~\cite{Palenzuela:2007HeadOn}. Phase dependence also appears
in orbital binaries and more recent binary
studies~\cite{Palenzuela:2008Orbital,Siemonsen:2023hko,Siemonsen:2023age}.
Related work has examined scalar-soliton
collisions~\cite{Helfer:2018vtq}, highly compact boson-star
mergers~\cite{Palenzuela:2017kcg,Bezares:2022obu}, boson-star inspirals
and merger
signals~\cite{Evstafyeva:2022bpr,Evstafyeva:2024qvp,Evstafyeva:2026juq},
head-on mergers~\cite{Ge:2024fum,Ge:2025btw}, relativistic
scattering~\cite{Damour:2025oys}, and mixed black-hole--boson-star
systems~\cite{Marks:2026xvo,Ning:2026qxs}.

Our question is whether the relative phase mainly affects the initial
formation of the common bridge or its subsequent survival after contact.
These are distinct possibilities: a bridge may form for all phases but
remain coherent to different degrees during the subsequent nonlinear
evolution.

For radially excited boson stars, our previous work identified a
separate control on chain formation~\cite{Ge:2026wzh}. Visible
chain-like transients appeared only when the binary collision occurred
at a stage compatible with the breathing dynamics of the corresponding
isolated star. Varying the initial separation shifted the collision
relative to the same internal clock and produced the corresponding
turn-on and turn-off of the chain. Collision timing therefore controls
whether the encounter occurs at a stage favourable to visible
chain-like morphology.

\begin{figure}
    \centering
    \begin{minipage}[t]{0.49\textwidth}
        \centering
        \includegraphics[width=\linewidth]
        {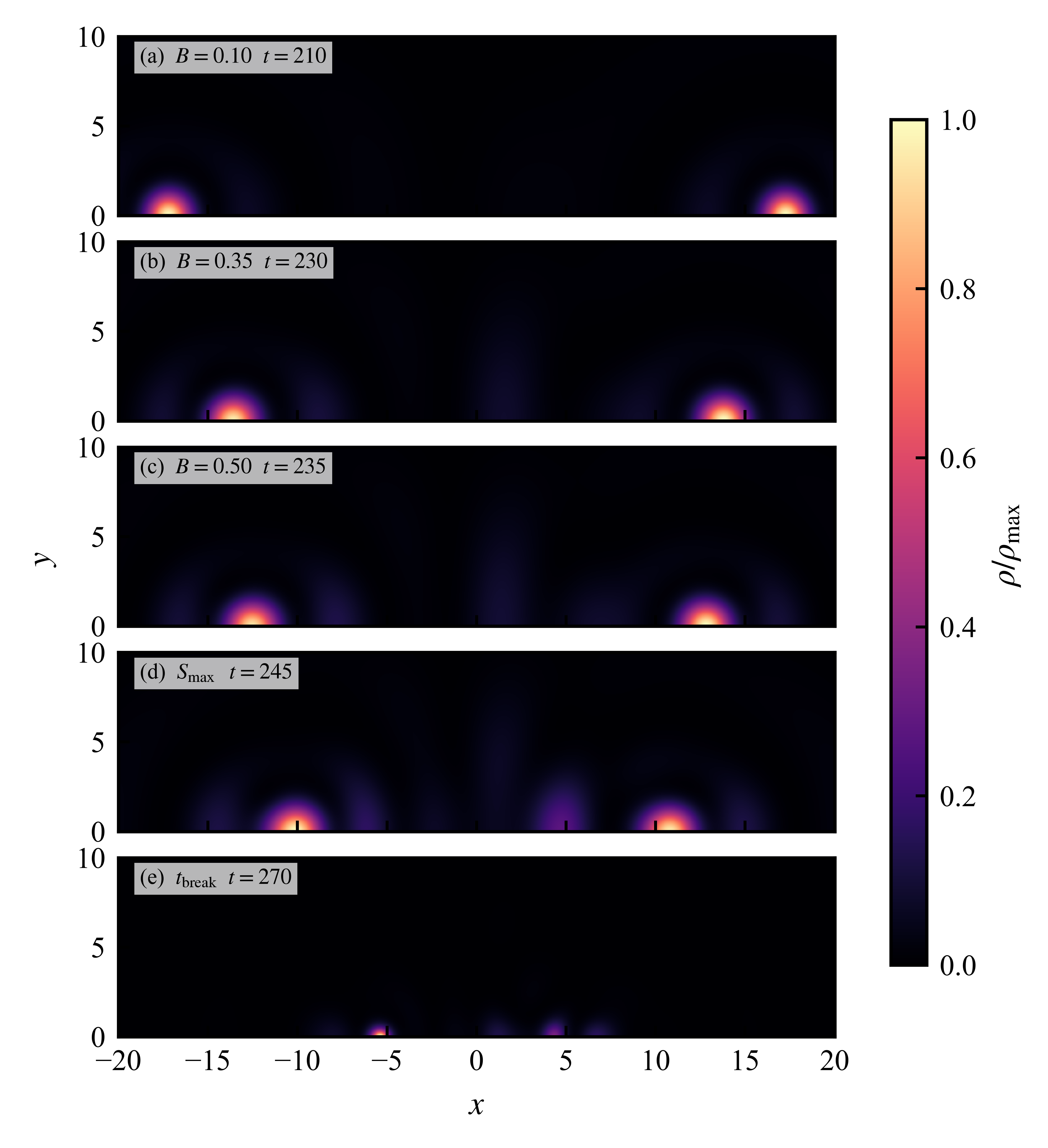}
    \end{minipage}
    \hfill
    \begin{minipage}[t]{0.49\textwidth}
        \centering
        \includegraphics[width=\linewidth]
        {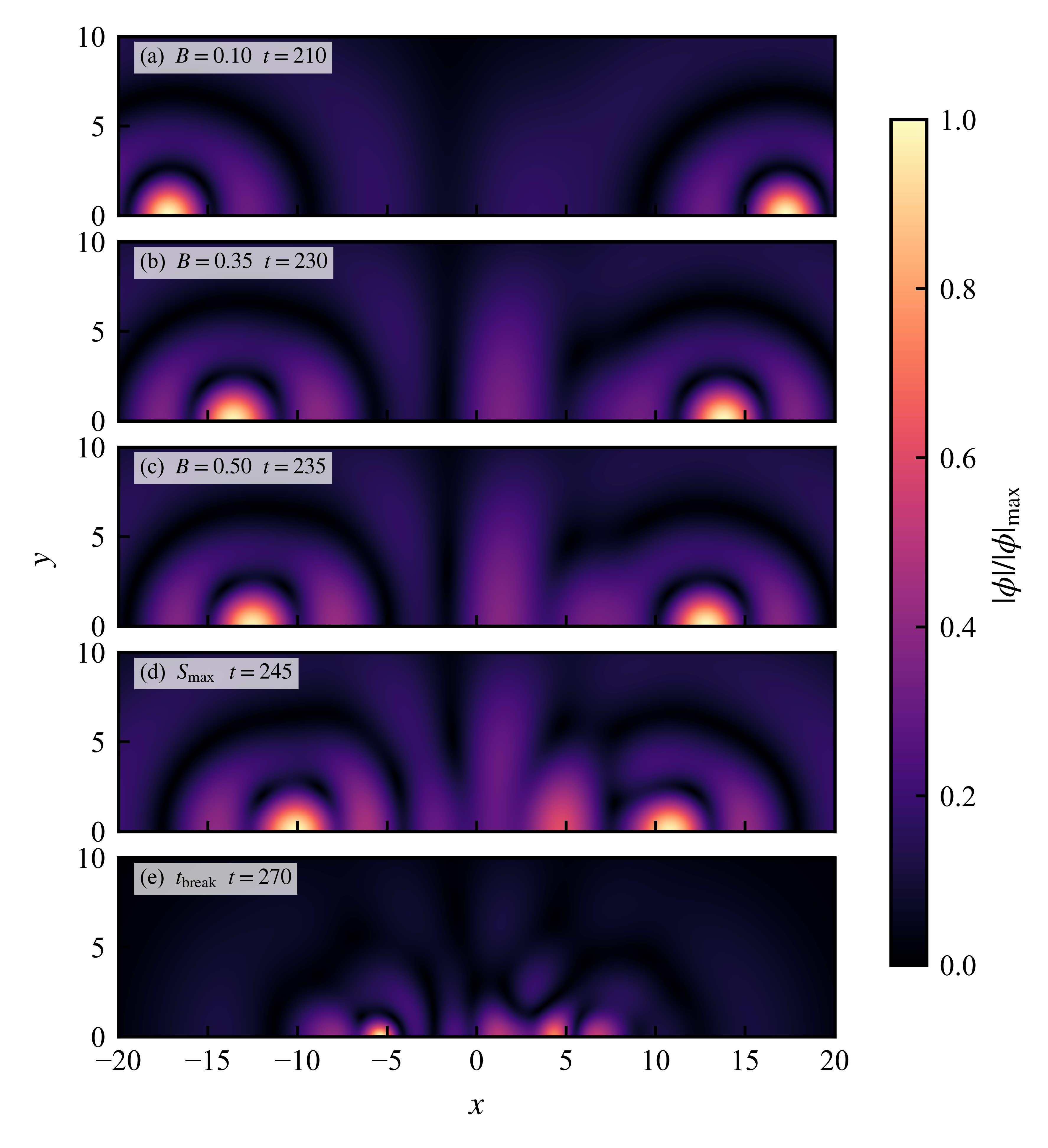}
    \end{minipage}
\caption{
Representative common-bridge evolution for the benchmark
\(\Delta\theta=\pi/2\) case.
The left and right columns show the normalised energy density
\(\rho/\rho_{\max}\) and scalar-field magnitude
\(|\phi|/|\phi|_{\max}\), respectively, at the same stages of the
evolution.
Here \(\rho_{\max}(t)\) and \(|\phi|_{\max}(t)\) denote the spatial
maxima at each output time, so that each snapshot is normalised
independently.
From top to bottom, the snapshots are the nearest available outputs to
the \(B=0.10\), \(B=0.35\), and \(B=0.50\) events, followed by the
nearest output to the time \(t(S_{\max})\) at which
\(S_{\rm chain}\) reaches its maximum \(S_{\max}\), and the first
sustained bridge breakup \(t_{\rm break}\), as labelled.
The energy density emphasises the morphology, while the scalar-field
magnitude makes the common bridge more visible. The event definitions
are given in Sec.~\ref{sec:setup}.
}
\label{fig:intro_bridge}
\end{figure}

Within the same collision setting, the present work separates
common-bridge formation from its subsequent survival.
Figure~\ref{fig:intro_bridge} illustrates a representative
common-bridge evolution through the energy density and the scalar-field
magnitude. The former emphasises the density morphology, while the
latter makes the developing bridge more apparent. The central region
first fills as the stars approach, develops a finite-amplitude common
bridge, and can later lose that connectivity. The chain score
\(S_{\rm chain}\), introduced in Ref.~\cite{Ge:2026wzh}, characterises
the instantaneous multi-peak morphology of this evolving configuration,
rather than the connectivity of the bridge itself. Its maximum value
therefore identifies the most pronounced chain-like morphology, but does
not determine how long the common bridge remains connected.

We study this question by varying the relative phase
\(\Delta\theta\) in equal-mass head-on collisions of radially excited
\(n=2\) boson stars. The \(n=2\) configurations and the head-on setup
retain direct continuity with Ref.~\cite{Ge:2026wzh}, allowing the
relative phase to be introduced as a separate control parameter within
the same class of collision problem.

In the benchmark configuration, all five sampled phases reach both the
common- and strong-bridge levels. Once the evolutions are aligned at the
operational contact time \(t_{\rm contact}\), their early post-contact
evolution remains similar; the phase dependence becomes clear only
later. Over the interval \(30\leq\tau\leq50\), the anti-phase case
retains about \(63\%\) of the accumulated bridge connectivity of the
in-phase case. The delayed separation after the same weak-contact level
has been reached shows that the phase dependence is not exhausted by an
instantaneous reduction of the scalar-field overlap.

The same phase-ordered suppression persists when the central scalar
amplitude and quartic self-interaction strength are varied, and is also
present for solitonic boson stars. Relative phase therefore acts
primarily on the persistence of an already formed common bridge. We
discuss this behaviour in terms of the spatial phase variation required
to connect finite-amplitude scalar configurations with a nonzero phase
mismatch.

Together with Ref.~\cite{Ge:2026wzh}, the two studies identify distinct
controls on the nonlinear dynamics. Collision timing governs whether
the encounter occurs at a stage favourable to visible chain formation,
whereas relative phase governs the persistence of the common bridge once
a bridge has formed.

Throughout this work, we employ geometric units with \(G=c=1\) and set
the scalar-field mass to \(m=1\).

\section{Model, numerical setup, and bridge diagnostics}
\label{sec:setup}

\subsection{Boson-star models}

We consider a complex scalar field minimally coupled to gravity, with action
\begin{equation}
S=\int d^4x\,\sqrt{-g}\left[\frac{R}{16\pi}
-\frac{1}{2}g^{\mu\nu}\nabla_\mu\bar{\phi}\nabla_\nu\phi
-\frac{1}{2}V(|\phi|^2)\right].
\label{eq:action}
\end{equation}

Our main simulations use the quartically self-interacting massive model,
\begin{equation}
V_{\rm M}(|\phi|^2)=m^2|\phi|^2+\frac{\lambda}{2}|\phi|^4,
\label{eq:massive_potential}
\end{equation}
while the robustness analysis also includes solitonic boson stars with
\begin{equation}
V_{\rm S}(|\phi|^2)=m^2|\phi|^2
\left(1-\frac{2|\phi|^2}{\sigma^2}\right)^2.
\label{eq:solitonic_potential}
\end{equation}

Equilibrium boson stars are constructed with the harmonic
ansatz~\cite{Kaup:1968zz,Ruffini:1969qy},
\begin{equation}
\phi(t,r)=|\phi(r)|e^{i\omega t},
\label{eq:bs_ansatz}
\end{equation}
where \(\omega\) is the field frequency and
\(|\phi_c|\equiv|\phi(0)|\) is the central scalar-field amplitude.
Throughout this work we consider the radially excited \(n=2\) branch,
where \(n\) denotes the number of radial nodes. We retain the \(n=2\)
configurations studied in Ref.~\cite{Ge:2026wzh}, so that the relative
phase can be introduced as a separate control parameter within a
previously characterised collision setting. The boson-star construction
and numerical framework follow our previous
work~\cite{Ge:2024fum,Ge:2024itl,Ge:2025btw,Ge:2026wzh}.

\subsection{Binary setup}

The numerical evolutions use the same {\sc GRChombo}-based
modified-Cartoon framework~\cite{Andrade:2021rbd}, AMR hierarchy, and
resolution as in Refs.~\cite{Ge:2024fum,Ge:2025btw,Ge:2026wzh}, with
the modified-Cartoon implementation following
Refs.~\cite{Cook:2016qnt,Cook:2016soy,Alcubierre:1999ab}.
Binary initial data are constructed from two equal-mass, nonspinning
boson stars using the improved superposition prescription of
Refs.~\cite{Helfer:2018vtq,Helfer:2021brt,Evstafyeva:2022bpr}.

We use head-on collisions as a controlled setting in which the phase
dependence can be isolated without additional effects from orbital
angular momentum and orbital evolution. The stars are initially
separated by \(D=80\). Unless stated otherwise, each star is given an
inward velocity \(v=0.1\). Our benchmark configuration is
\begin{equation}
\lambda=100,\qquad n=2,\qquad |\phi_c|=0.08,\qquad
D=80,\qquad v=0.1.
\label{eq:benchmark}
\end{equation}

To introduce a relative phase, we denote the scalar field and momentum
variable of the two individual stars entering the superposition by
\((\phi^{(1)},\Pi^{(1)})\) and
\((\phi^{(2)},\Pi^{(2)})\), respectively. Taking star \(1\) as the
phase reference, we apply a global \(U(1)\) rotation to star \(2\),
\begin{equation}
\phi^{(2)}\rightarrow e^{i\Delta\theta}\phi^{(2)},\qquad
\Pi^{(2)}\rightarrow e^{i\Delta\theta}\Pi^{(2)}.
\label{eq:phase_rotation}
\end{equation}
For the equal-mass head-on configurations considered here,
\(\Delta\theta\) and \(2\pi-\Delta\theta\) are related by interchange
of the two stars, so it is sufficient to restrict the scan to
\(0\leq\Delta\theta\leq\pi\).
Thus we sample five relative phases,
\begin{equation}
\Delta\theta=0,\qquad \frac{\pi}{4},\qquad \frac{\pi}{2},\qquad
\frac{3\pi}{4},\qquad \pi.
\label{eq:phase_values}
\end{equation}
All other initial parameters are held fixed within each phase scan.

\subsection{Common-bridge diagnostics}

We characterise the connectivity between the two stars using the bridge
ratio \(B(t)\), introduced in Ref.~\cite{Ge:2026wzh}. It compares the
near-axis density weight in the central region with the weaker of the
two stellar-side density peaks. Denoting these quantities by
\(P_{\rm br}\), \(P_L\), and \(P_R\), respectively, we write
\begin{equation}
B(t)=\frac{P_{\rm br}(t)}{\min[P_L(t),P_R(t)]}.
\label{eq:bridge_ratio}
\end{equation}
Thus \(B\ll1\) corresponds to two well-separated density
concentrations, while increasing \(B\) indicates progressive filling of
the region between them. The detailed construction of
\(P_{\rm br}\), \(P_L\), and \(P_R\) is given in
Appendix~\ref{app:bridge}.

For the timing analysis we use three reference levels,
\begin{equation}
B_{\rm contact}=0.10,\qquad
B_{\rm common}=0.35,\qquad
B_{\rm strong}=0.50.
\label{eq:bridge_thresholds}
\end{equation}
The corresponding event times are denoted by
\(t_{\rm contact}\), \(t_{\rm common}\), and \(t_{\rm strong}\).
These are operational markers for weak initial contact,
common-bridge formation, and strong-bridge formation, respectively.
Once a strong bridge has formed,
\(t_{\rm break}\) denotes its first sustained return below
\(B_{\rm common}\). The precise sustainment criteria are given in
Appendix~\ref{app:bridge}.

To separate differences in the operational contact time from the
subsequent bridge evolution, we define the contact-aligned time
\begin{equation}
\tau=t-t_{\rm contact}.
\label{eq:contact_time}
\end{equation}
Hereafter, \(B(\tau;\Delta\theta)\) denotes the same bridge ratio
expressed in this shifted time coordinate. We quantify the accumulated
bridge connectivity over an interval \([\tau_1,\tau_2]\) by
\begin{equation}
J_B(\tau_1,\tau_2;\Delta\theta)
=\int_{\tau_1}^{\tau_2}B(\tau;\Delta\theta)\,d\tau.
\label{eq:JB}
\end{equation}
For comparisons between different equilibrium configurations, we also
use the response normalised to the corresponding in-phase evolution,
\begin{equation}
\widehat{J}_B(\tau_1,\tau_2;\Delta\theta)
=\frac{J_B(\tau_1,\tau_2;\Delta\theta)}
{J_B(\tau_1,\tau_2;0)}.
\label{eq:JB_normalized}
\end{equation}

We retain the chain score \(S_{\rm chain}\) of
Ref.~\cite{Ge:2026wzh} as an auxiliary morphology diagnostic that
quantifies the prominence of the instantaneous chain-like morphology
in the energy-density distribution, and define
\(S_{\max}\equiv\max_t S_{\rm chain}(t)\).

The corresponding time \(t(S_{\max})\) identifies the most pronounced
chain-like configuration. We do not use \(S_{\max}\) or the number of
chain episodes as the primary measure of bridge survival: these
characterise instantaneous morphology, whereas \(J_B\) provides a
continuous measure of the accumulated bridge connectivity over a
specified time interval.

\section{Relative phase and the survival of the common bridge}
\label{sec:phase}

\begin{figure}
    \centering
    \includegraphics[width=0.75\textwidth]
    {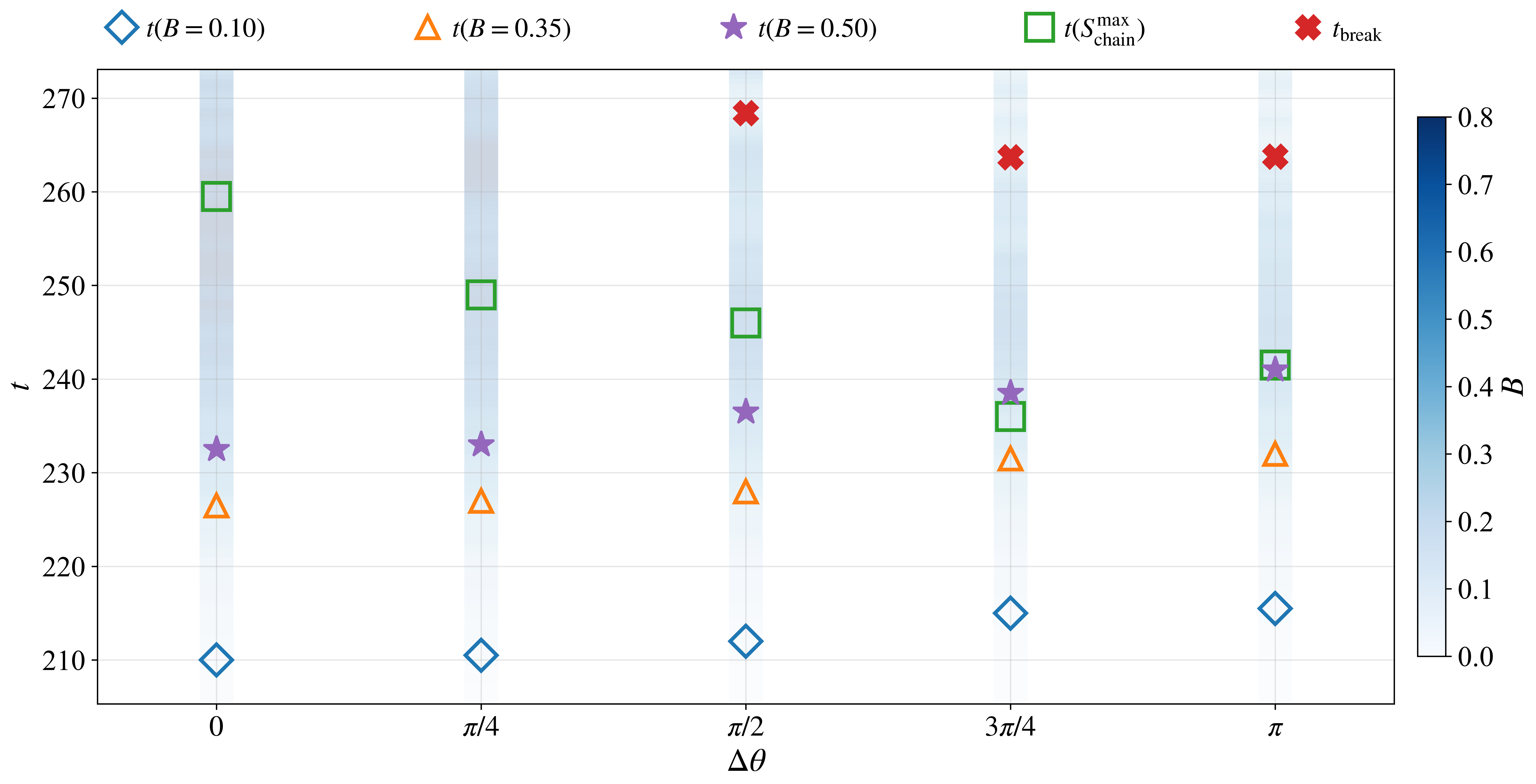}
    \caption{
    Characteristic times in the benchmark phase scan. The vertical
    coloured bands show the bridge ratio \(B(t)\), using the scale at
    right. The markers show the first sustained events associated with
    \(B=0.10\), \(0.35\), and \(0.50\), together with \(t(S_{\max})\)
    and the first sustained bridge breakup \(t_{\rm break}\).
    }
    \label{fig:benchmark_timeline}
\end{figure}

We first vary only the relative phase in the benchmark configuration of
Eq.~\eqref{eq:benchmark}. Figure~\ref{fig:benchmark_timeline} summarises
the characteristic times for the five phase offsets. Increasing
\(\Delta\theta\) produces a modest but systematic delay:
\(t_{\rm contact}\) shifts from \(210.0\) at \(\Delta\theta=0\) to
\(215.5\) at \(\Delta\theta=\pi\), while \(t_{\rm common}\) shifts from
\(226.5\) to \(232.0\). All five cases nevertheless reach both the
common- and strong-bridge thresholds. The relative phase therefore does
not prevent the initial formation of a common bridge in this
configuration.

\begin{figure}
    \centering
    \includegraphics[width=0.485\textwidth]
    {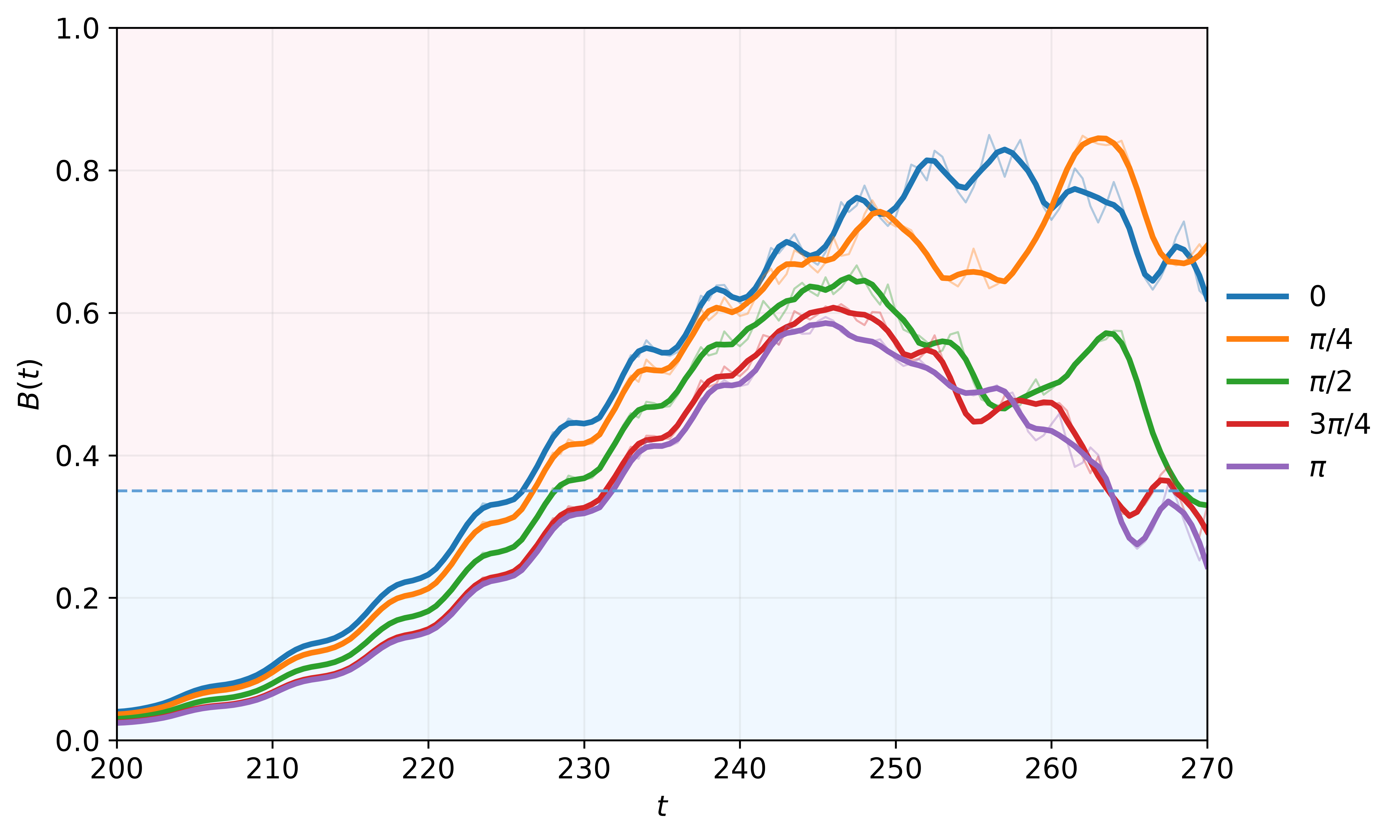}
    \hfill
    \includegraphics[width=0.485\textwidth]
    {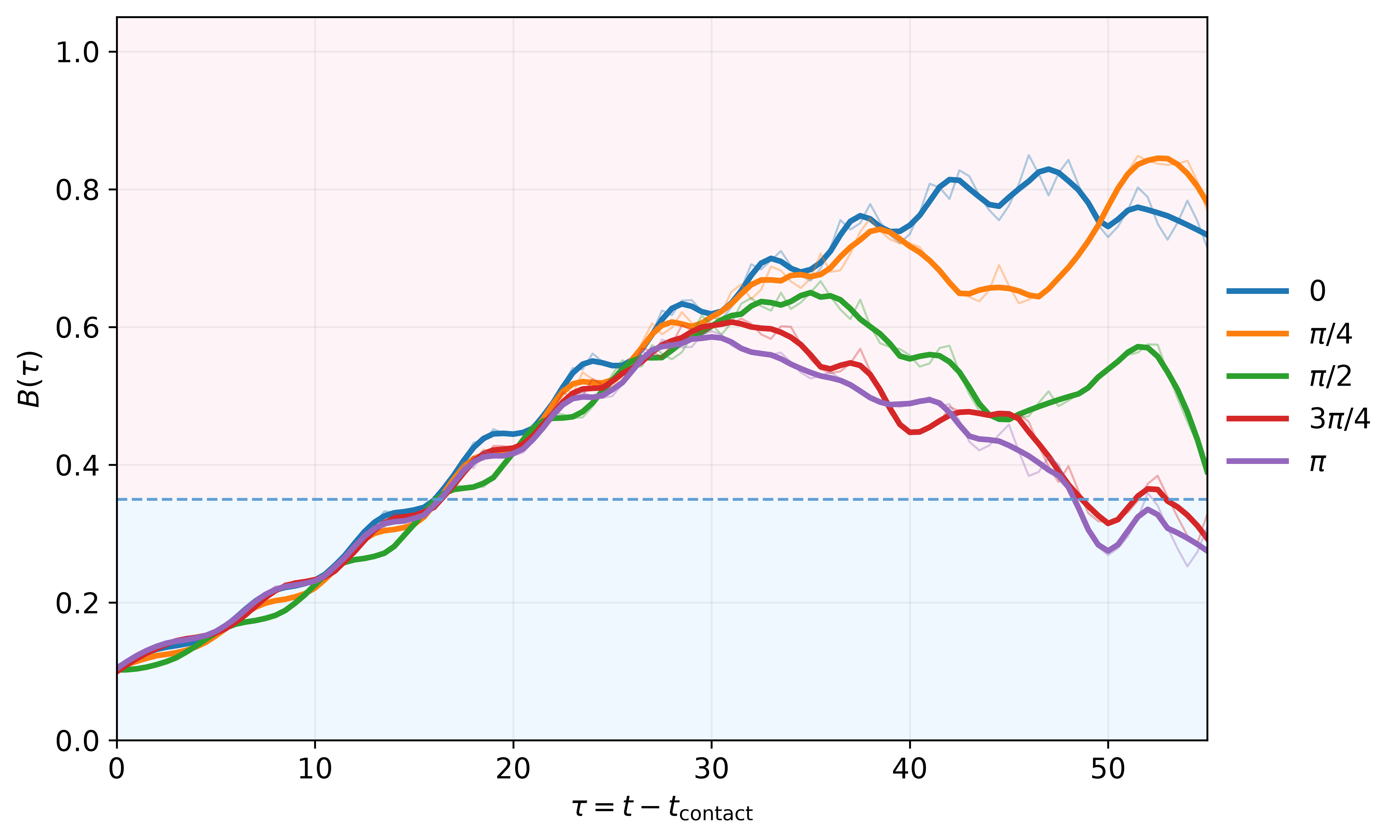}
    \caption{
    Bridge evolution for the five relative phases of the benchmark
    configuration. Left: bridge ratio \(B(t)\) in the original evolution
    time. Right: the same evolutions aligned at
    \(t_{\rm contact}\), with \(\tau=t-t_{\rm contact}\).
    Faint curves show the raw diagnostic data, while the thicker curves
    are smoothed guides to the eye; all quantitative results use the raw
    data. The dashed line marks \(B_{\rm common}=0.35\), with the shaded
    regions indicating values below and above this level. The initial
    post-contact build-up is similar across the phase scan, while the
    subsequent bridge evolution separates with increasing relative
    phase.
    }
    \label{fig:benchmark_bridge}
\end{figure}

The later evolution is more strongly phase dependent.
Figure~\ref{fig:benchmark_bridge} (left) shows the bridge ratio
\(B(t)\) directly. The smaller-offset cases develop a stronger and more
persistent central connection, whereas the larger-offset cases reach
lower bridge ratios and weaken earlier. Sustained breakup occurs for
\(\Delta\theta=\pi/2\), \(3\pi/4\), and \(\pi\), while the two
smaller-offset cases remain strongly connected over the interval shown.
The main phase dependence is therefore associated with the persistence
of the bridge rather than with its initial formation.

This distinction becomes clearer after removing the phase-dependent
shift in \(t_{\rm contact}\). Figure~\ref{fig:benchmark_bridge} (right)
shows the same evolutions in the contact-aligned coordinate
\(\tau=t-t_{\rm contact}\). During the first \(\sim20\) time units
after this marker, the five bridge ratios remain close, showing
comparable initial bridge build-up. A clear separation develops only
later, beginning around \(\tau\simeq25\)--\(30\). The smaller-offset
systems remain strongly connected, whereas the larger-offset systems
turn over and progressively lose connectivity.

Motivated by this separation, we use
\begin{equation}
30\leq\tau\leq50
\label{eq:bridge_survival_window}
\end{equation}
as a common post-contact window for the quantitative comparisons below.
This interval lies after the initial bridge build-up and covers the
stage where the phase-dependent loss of connectivity is pronounced. For
the benchmark configuration, direct integration of \(B(\tau)\) gives
\begin{equation}
J_B(30,50;\Delta\theta)
=
\{14.958,\ 13.632,\ 11.269,\ 9.877,\ 9.485\},
\qquad
\Delta\theta=
\left\{0,\frac{\pi}{4},\frac{\pi}{2},\frac{3\pi}{4},\pi\right\}.
\label{eq:benchmark_JB_absolute}
\end{equation}
Normalising to the in-phase evolution yields
\begin{equation}
\widehat{J}_B(30,50;\Delta\theta)
=
\{1.000,\ 0.911,\ 0.753,\ 0.660,\ 0.634\}.
\label{eq:benchmark_JB}
\end{equation}
Thus the anti-phase case retains about \(63\%\) of the accumulated
post-contact bridge connectivity of the in-phase case over the same
interval.

The delayed separation of the contact-aligned evolutions shows that the
observed phase dependence cannot be characterised solely as an
instantaneous reduction of the scalar-field overlap at contact. The
benchmark instead points to a dynamical effect that develops after the
common bridge has formed: relative phase primarily controls its
post-contact survival rather than its initial formation. In the next
section we test whether this behaviour persists when the equilibrium
configuration and scalar potential are changed.

\section{Robustness of the phase response}
\label{sec:robustness}

We next test whether the phase response found in the benchmark persists
away from the fiducial configuration. These scans are controlled
robustness tests rather than an exhaustive survey of the boson-star
parameter space. In each comparison, one aspect of the configuration is
changed while the remaining parameters are kept fixed. The same five
relative phases are used throughout, with \(J_B\) evaluated over the
bridge-survival window \(30\leq\tau\leq50\).

\subsection{Central-amplitude dependence}

\begin{figure}
    \centering
    \includegraphics[width=0.485\textwidth]
    {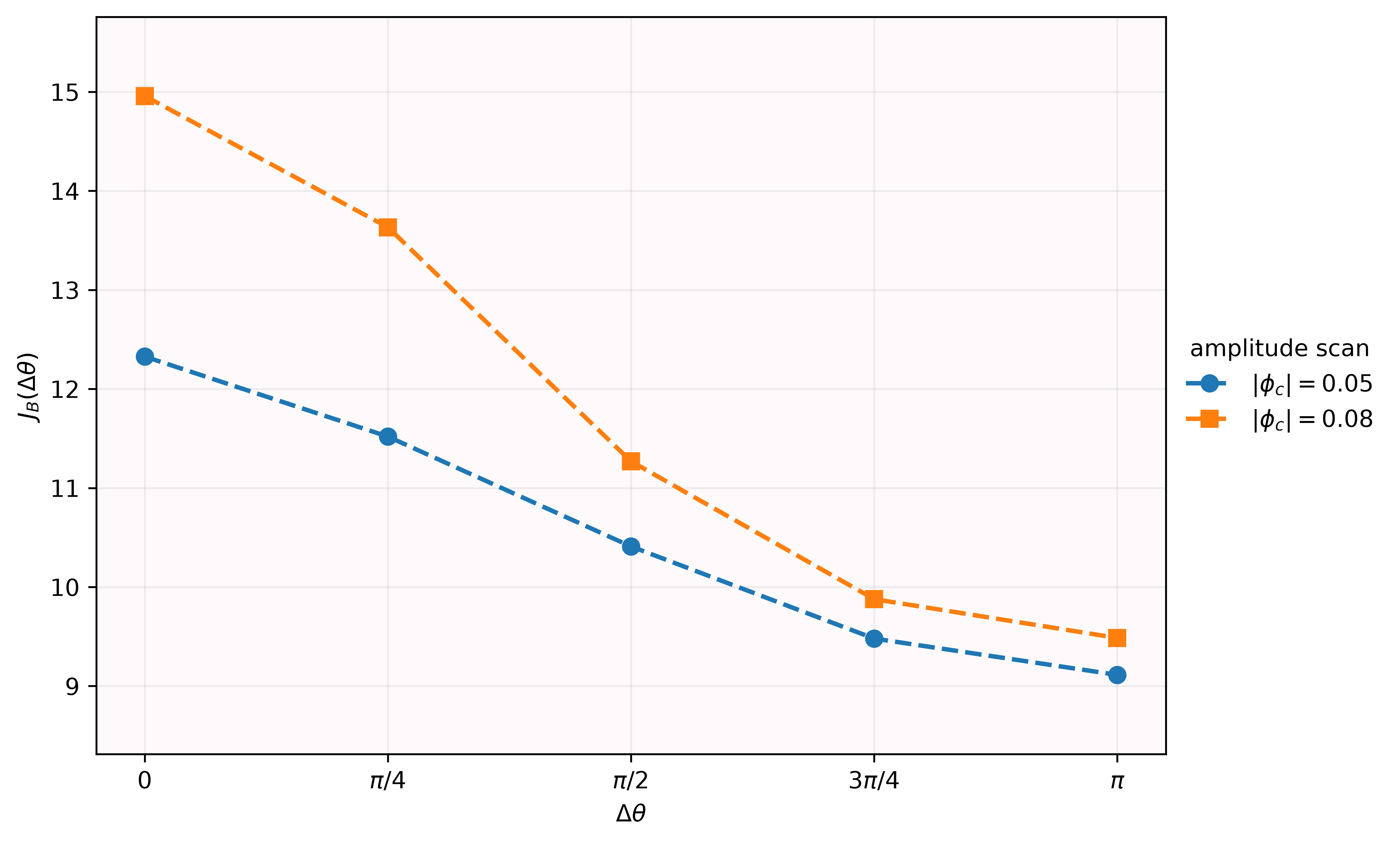}
    \hfill
    \includegraphics[width=0.485\textwidth]
    {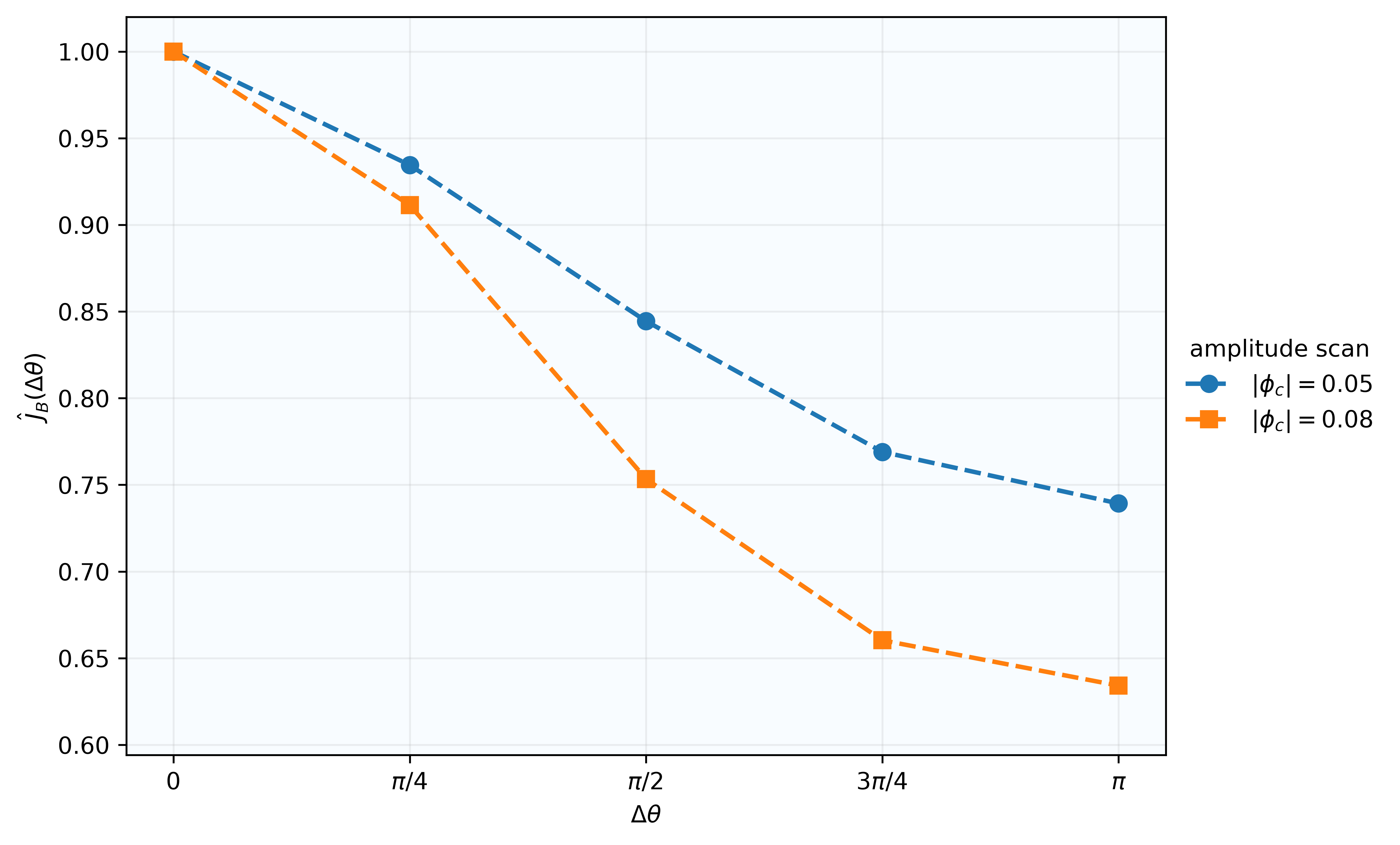}
    \caption{
    Post-contact bridge response for
    \(|\phi_c|=0.05\) and \(0.08\) at fixed
    \(\lambda=100\) and \(v=0.1\).
    Left: \(J_B(30,50;\Delta\theta)\).
    Right: the same quantity normalised to the corresponding in-phase
    evolution.
    }
    \label{fig:amplitude_response}
\end{figure}

We first keep \(\lambda=100\) and \(v=0.1\) fixed and compare
\(|\phi_c|=0.05\) with the benchmark value \(|\phi_c|=0.08\).
Figure~\ref{fig:amplitude_response} shows \(J_B\) and its normalised
counterpart \(\widehat J_B\). The benchmark has a larger \(J_B\)
throughout the phase scan, but both amplitudes show the same overall
phase ordering.

At \(\Delta\theta=\pi\),
\(\widehat J_B=0.739\) for \(|\phi_c|=0.05\) and
\(\widehat J_B=0.634\) for \(|\phi_c|=0.08\).
For both configurations shown here, \(J_B\) decreases monotonically
across the sampled phases, although the strength of the response depends
on \(|\phi_c|\).

\subsection{Quartic self-interaction}

\begin{figure}
    \centering
    \includegraphics[width=0.485\textwidth]
    {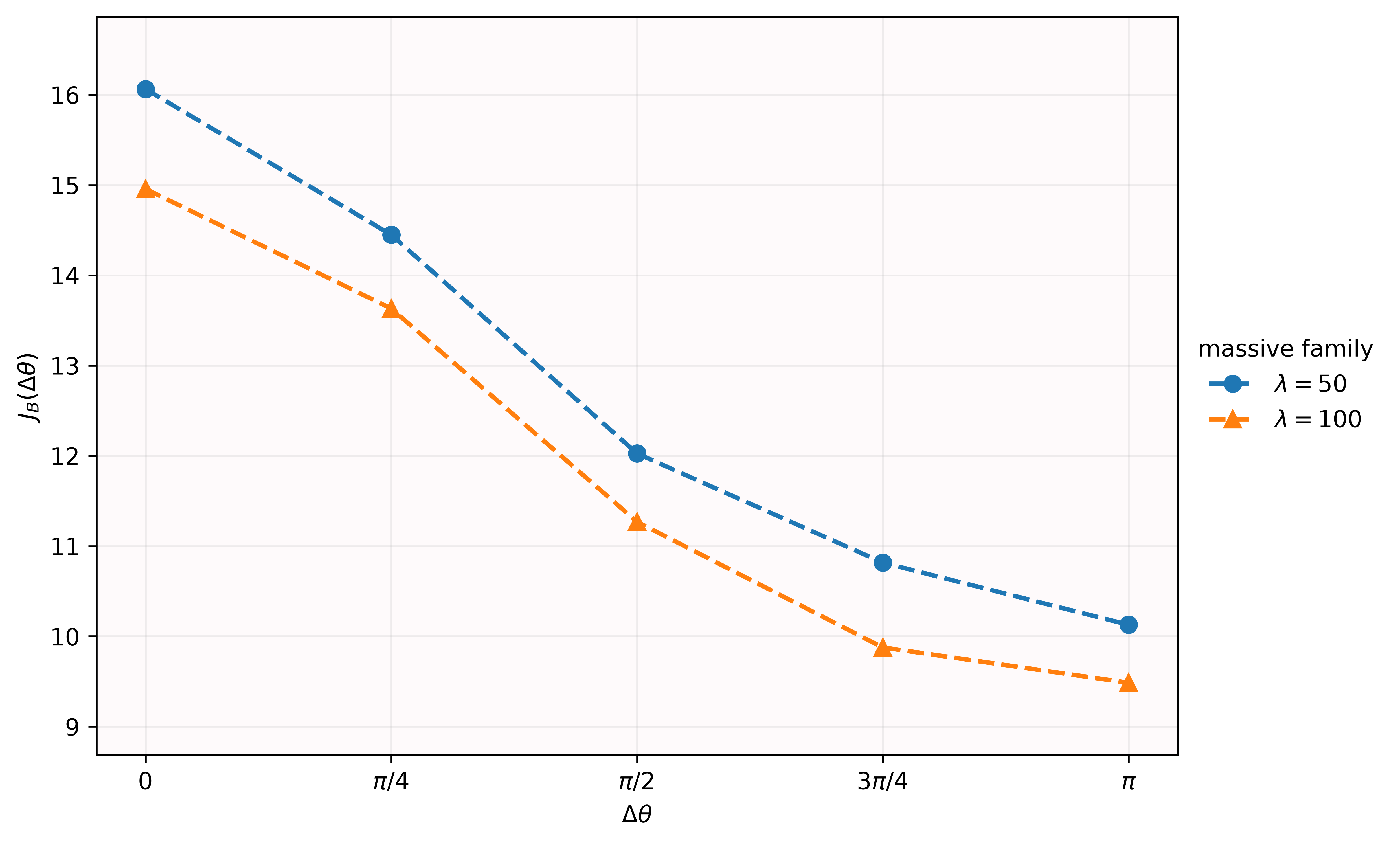}
    \hfill
    \includegraphics[width=0.485\textwidth]
    {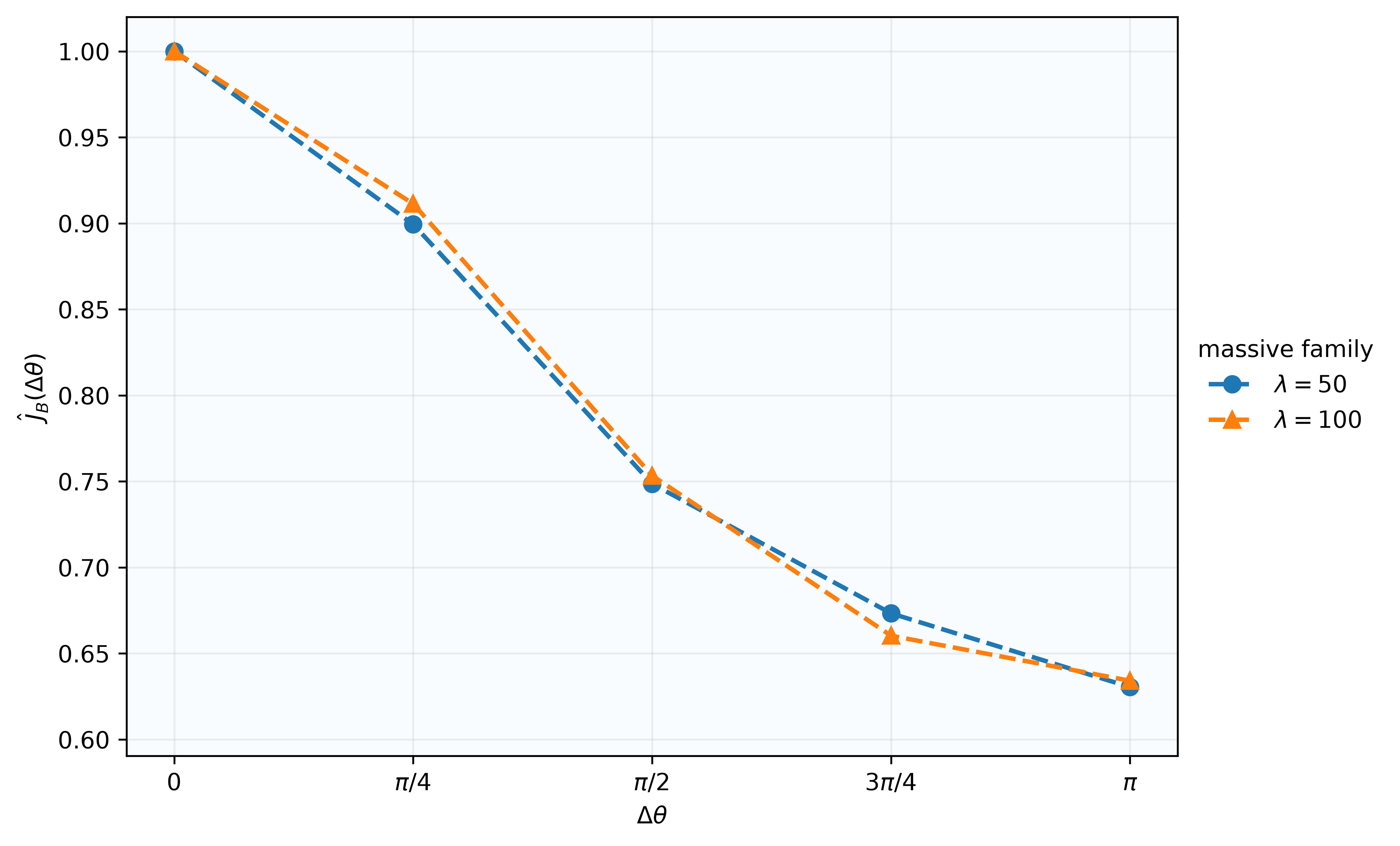}
    \caption{
    Post-contact bridge response for
    \(\lambda=50\) and \(100\) at fixed
    \(|\phi_c|=0.08\) and \(v=0.1\).
    Left: \(J_B(30,50;\Delta\theta)\).
    Right: the response normalised to the corresponding in-phase
    evolution.
    }
    \label{fig:lambda_response}
\end{figure}

We next fix \(|\phi_c|=0.08\) and \(v=0.1\), and compare
\(\lambda=50\) with the benchmark value \(\lambda=100\).
The \(\lambda=50\) configuration has a larger \(J_B\) throughout the
phase scan, while the normalised curves in
Fig.~\ref{fig:lambda_response} are nearly coincident.

At \(\Delta\theta=\pi\),
\(\widehat J_B=0.630\) for \(\lambda=50\) and
\(0.634\) for \(\lambda=100\). The anti-phase evolution therefore
retains about \(63\%\) of the corresponding in-phase accumulated
connectivity in both cases. The normalised phase response is nearly
unchanged under this factor-of-two variation in \(\lambda\), despite
the change in the absolute scale of \(J_B\).

\subsection{Solitonic boson stars}

\begin{figure}
    \centering
    \includegraphics[width=0.485\textwidth]
    {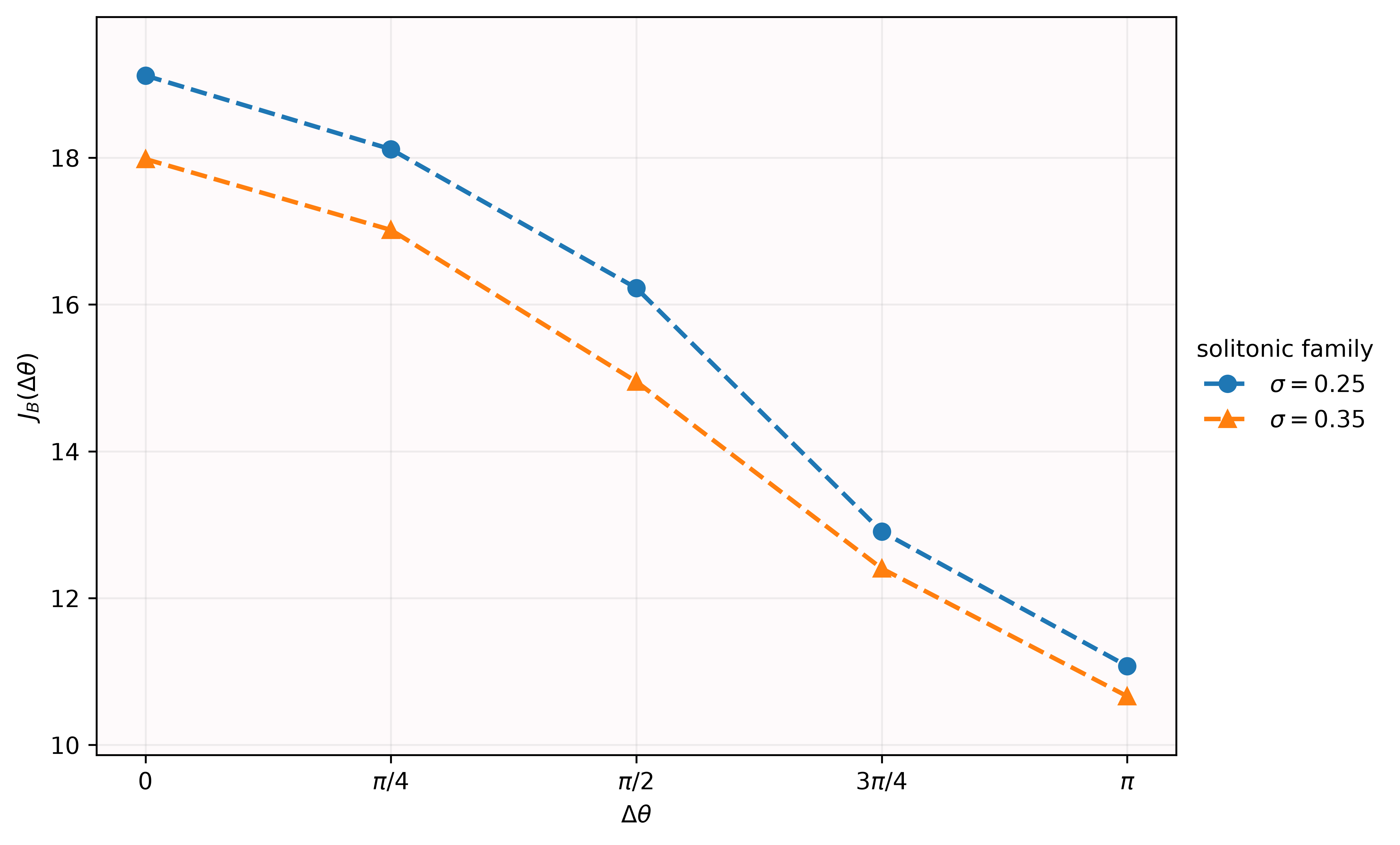}
    \hfill
    \includegraphics[width=0.485\textwidth]
    {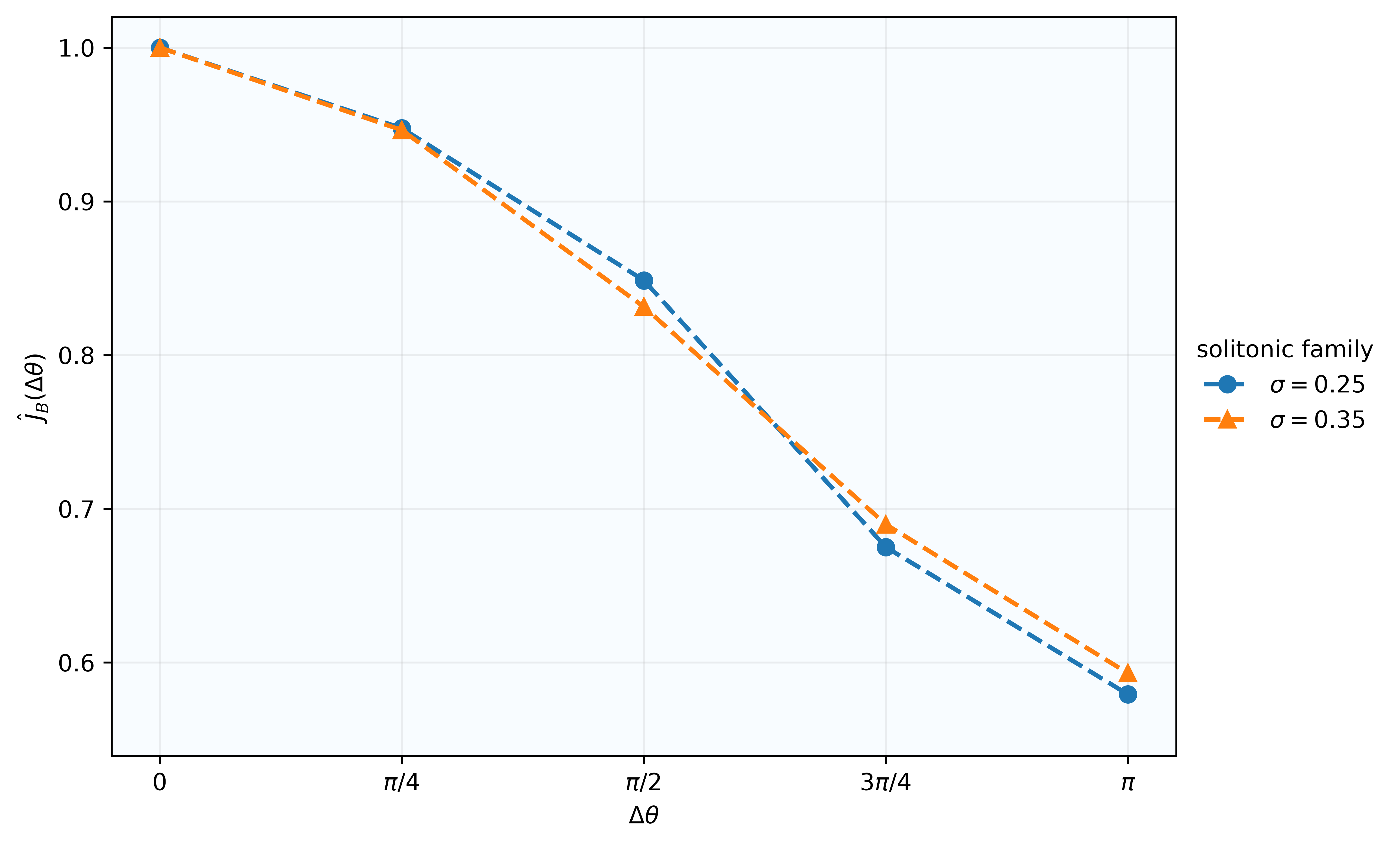}
    \caption{
    Post-contact bridge response for solitonic boson stars with
    \(\sigma=0.25\) and \(0.35\), at fixed
    \(|\phi_c|=0.08\) and \(v=0.1\).
    Left: \(J_B(30,50;\Delta\theta)\).
    Right: the response normalised to the corresponding in-phase
    evolution.
    }
    \label{fig:sigma_response}
\end{figure}

We finally change the scalar potential itself. For the solitonic model
of Eq.~\eqref{eq:solitonic_potential}, we take
\(\sigma=0.25\) and \(0.35\), keeping
\(|\phi_c|=0.08\) and \(v=0.1\).
The results are shown in Fig.~\ref{fig:sigma_response}.

The unnormalised \(J_B\) again depends on the equilibrium configuration,
whereas the normalised curves remain close over the full phase range.
At \(\Delta\theta=\pi\), the two configurations retain about
\(58\)--\(59\%\) of their corresponding in-phase accumulated
connectivity. Both solitonic scans decrease monotonically across the
five sampled phases.

Across these scans, the absolute scale of \(J_B\) remains
configuration dependent. For all configurations shown here, however,
\(J_B\) decreases monotonically across the five sampled phases.
The benchmark suppression is therefore not tied to a single central
amplitude, quartic coupling, or scalar potential.

\section{Physical interpretation}
\label{sec:interpretation}

The contact-aligned results constrain the interpretation of the phase
dependence. Increasing \(\Delta\theta\) produces a modest delay in
\(t_{\rm contact}\), but this shift alone cannot explain the subsequent
bridge hierarchy. After alignment at \(t_{\rm contact}\), the five
bridge ratios remain close for roughly the first \(20\) time units and
begin to separate only around \(\tau\simeq25\)--\(30\); see
Fig.~\ref{fig:benchmark_bridge}. The main dynamical effect of the
relative phase therefore emerges during the post-contact evolution.

At the level of the superposed scalar field, the phase dependence
appears through
\begin{equation}
|\phi_1+e^{i\Delta\theta}\phi_2|^2
=
|\phi_1|^2+|\phi_2|^2
+2\,{\rm Re}\!\left(e^{i\Delta\theta}\bar{\phi}_1\phi_2\right).
\label{eq:phase_interference}
\end{equation}
The relative phase can therefore enhance or suppress the local field
magnitude in the overlap region. This algebraic effect alone does not
account for the delayed bridge hierarchy: all five benchmark phases
reach both the common- and strong-bridge thresholds, and their
contact-aligned bridge ratios remain close during the early
post-contact stage.

The subsequent evolution suggests a phase-gradient interpretation.
Extending the notation of Eq.~\eqref{eq:bs_ansatz}, we write the
evolving field locally as
\(\phi=|\phi|e^{i\vartheta}\), where \(\vartheta\) is the local phase.
On a spatial slice,
\begin{equation}
\gamma^{ij}\partial_i\bar{\phi}\,\partial_j\phi
=
\gamma^{ij}\partial_i|\phi|\,\partial_j|\phi|
+
|\phi|^2\gamma^{ij}\partial_i\vartheta\,\partial_j\vartheta,
\label{eq:phase_gradient}
\end{equation}
where \(\gamma^{ij}\) is the inverse spatial metric. For the in-phase
case, no phase mismatch is imposed between the two sides by the initial
data. At nonzero \(\Delta\theta\), a persistent finite-amplitude
connection must accommodate the mismatch through spatial phase
variation across the bridge. The corresponding gradient contribution
is weighted by \(|\phi|^2\), and can therefore be reduced if the field
amplitude decreases where the phase changes rapidly. A large phase
offset can thus disfavour a broad, persistent high-amplitude connection.

The amplitude scan is also consistent with this interpretation.
At \(\Delta\theta=\pi\), the normalised response is
\(\widehat J_B\simeq0.739\) for \(|\phi_c|=0.05\), compared with
\(\widehat J_B\simeq0.634\) for the benchmark
\(|\phi_c|=0.08\); see Fig.~\ref{fig:amplitude_response}. The
lower-amplitude configuration is therefore less sensitive to the phase
offset, consistent with the \(|\phi|^2\) weighting in
Eq.~\eqref{eq:phase_gradient}. Since changing \(|\phi_c|\) also changes
the underlying equilibrium star, this comparison supports rather than
directly tests the phase-gradient interpretation.

The same picture is consistent with the robustness under changes of the
scalar potential. The massive and solitonic models have different
self-interaction potentials, but both are \(U(1)\)-symmetric and share
the same canonical phase-gradient structure. Their absolute bridge
responses differ, while the suppression with increasing
\(\Delta\theta\) persists; see Figs.~\ref{fig:lambda_response} and
\ref{fig:sigma_response}.

This interpretation is complementary to the timing-window mechanism of
Ref.~\cite{Ge:2026wzh}. There the collision time controls whether the
binary encounters a stage compatible with visible chain formation. Here
all five benchmark phases form a strong bridge, while the survival
hierarchy remains after alignment at \(t_{\rm contact}\). Relative phase
therefore acts primarily on the persistence of an already formed common
bridge. We do not separately measure the bridge-region phase-gradient
contribution here, so Eq.~\eqref{eq:phase_gradient} provides a physical
interpretation rather than a direct diagnostic.

\section{Conclusion}
\label{sec:conclusion}

We have studied whether the relative phase primarily affects the
formation of a common bridge or its subsequent post-contact survival in
head-on collisions of radially excited \(n=2\) boson stars. We
distinguish common-bridge connectivity from the instantaneous
chain-like morphology. The common bridge is a more general
finite-amplitude connection whose definition does not rely on radial
excitation, whereas \(S_{\rm chain}\) characterises the multi-peak
morphology associated here with the excited-state configurations and is
used only as an auxiliary morphology diagnostic.

In the benchmark configuration, all five sampled phases reach both the
common- and strong-bridge thresholds. After aligning the evolutions at
\(t_{\rm contact}\), the bridge ratios remain close during the early
post-contact stage and separate only later. The delayed separation shows
that the observed phase dependence cannot be characterised solely as an
instantaneous reduction of the scalar-field overlap at contact. Relative
phase instead acts primarily on the survival of an already formed common
bridge rather than on its initial formation.

Over the post-contact interval \(30\leq\tau\leq50\), the anti-phase
benchmark retains about \(63\%\) of the accumulated bridge connectivity
of the in-phase case. The absolute response depends on the equilibrium
configuration, but the phase-ordered suppression persists across the
sampled central amplitudes, quartic couplings, and solitonic models. The
effect is therefore not tied to a single equilibrium configuration or
to one choice of scalar potential within the parameter ranges studied.

The results are consistent with a phase-gradient interpretation. A
persistent finite-amplitude connection between regions with different
phases must accommodate the mismatch through spatial phase variation,
whose gradient contribution is weighted by \(|\phi|^2\). Large phase
offsets can therefore disfavour a broad, persistent high-amplitude
connection. This argument provides a physical interpretation of the
observed survival hierarchy rather than a direct diagnostic of the
bridge-region phase gradient.

Together with the timing-window mechanism identified in our previous
work, these results separate two aspects of the excited-state collision
dynamics. Collision timing controls whether the encounter occurs at a
stage favourable to visible chain formation, while relative phase
controls how long the resulting common bridge remains coherent. In this
sense, chain formation and common-bridge survival are distinct dynamical
questions: the former is tied to the internal evolution of the excited
configuration, whereas the latter reflects the phase-dependent
persistence of the finite-amplitude connection formed during the
collision.

\begin{acknowledgments}
The author acknowledges HIAS for access to the ``Quantum Universe
Physical Simulation Platform''. This work was supported by the
National Natural Science Foundation of China under Grant No.~12505066.
\end{acknowledgments}

\clearpage
\appendix

\section{Bridge-diagnostic construction}
\label{app:bridge}

The bridge diagnostic is constructed from the energy density in the
axisymmetric sampling region. Let \(X_i\) denote the centre of the
\(i\)-th sampling interval along the collision axis, with spacing
\(\Delta X\). We define the near-axis density weight
\begin{equation}
P_x(X_i,t)
=
\int_{X_i-\Delta X/2}^{X_i+\Delta X/2} dx
\int_0^{R_{\rm ax}} dy\,
\rho(t,x,y)\sqrt{\gamma(t,x,y)}\,2\pi y,
\label{eq:app_Px}
\end{equation}
where \(\rho\) is the energy density and \(\gamma\) is the determinant
of the spatial metric. The factor \(2\pi y\) accounts for the
axisymmetric volume element. In practice, Eq.~\eqref{eq:app_Px} is
evaluated as a discrete sum over the diagnostic grid.

The transverse extent of the near-axis region is chosen as
\begin{equation}
R_{\rm ax}=0.1R_{\rm samp},
\label{eq:app_Rax}
\end{equation}
where \(R_{\rm samp}\) is the radius of the diagnostic sampling region.
For all runs considered in the main text we use
\begin{equation}
R_{\rm samp}=160,\qquad
N_x=320,\qquad
N_y=160,
\label{eq:app_sampling}
\end{equation}
which gives
\begin{equation}
\Delta X=1,\qquad R_{\rm ax}=16.
\label{eq:app_sampling_scales}
\end{equation}
Restricting the transverse integration to this near-axis region
suppresses contributions from off-axis material and makes the diagnostic
sensitive to the filling of the region between the two stars.

Let \(x_c\) be the midpoint of the initial binary and \(i_c\) the
sampling interval closest to it. The central bridge weight is obtained
by averaging over seven neighbouring axial intervals,
\begin{equation}
P_{\rm br}(t)
=
\frac{1}{2n_{\rm br}+1}
\sum_{k=-n_{\rm br}}^{n_{\rm br}}
P_x(X_{i_c+k},t),
\qquad
n_{\rm br}=3.
\label{eq:app_Pbridge}
\end{equation}
This averaging prevents a single sampling interval from dominating the
bridge estimate.

The left- and right-side peaks are extracted from the same near-axis
profile, excluding the central bridge region,
\begin{equation}
P_L(t)
=
\max_{i<i_c-n_{\rm br}} P_x(X_i,t),
\qquad
P_R(t)
=
\max_{i>i_c+n_{\rm br}} P_x(X_i,t).
\label{eq:app_PLPR}
\end{equation}
The bridge ratio used throughout the main text is then
\begin{equation}
B(t)
=
\frac{P_{\rm br}(t)}
{\min[P_L(t),P_R(t)]}.
\label{eq:app_bridge_ratio}
\end{equation}
Normalising by the weaker stellar-side peak reduces sensitivity to small
left--right asymmetries. For two separated configurations the central
region is depleted and \(B\ll1\); \(B\) increases as this region becomes
filled during the collision.

We use the three reference levels
\begin{equation}
B_{\rm contact}=0.10,\qquad
B_{\rm common}=0.35,\qquad
B_{\rm strong}=0.50.
\label{eq:app_bridge_levels}
\end{equation}
For each level \(B_\ast\), the corresponding event time is assigned to
the first output of the earliest sequence of three consecutive
diagnostic outputs satisfying
\begin{equation}
B(t)\geq B_\ast.
\label{eq:app_sustain_up}
\end{equation}
The sustainment condition removes isolated threshold crossings caused
by short-time fluctuations.

After the strong-bridge level has been reached, bridge breakup is
defined by the first downward crossing of \(B_{\rm common}\) for which
\begin{equation}
B(t)<B_{\rm common}
\label{eq:app_sustain_down}
\end{equation}
holds for three consecutive diagnostic outputs. We require an actual
downward crossing, with the preceding output satisfying
\(B\geq B_{\rm common}\). Once the sustainment condition is met,
\(t_{\rm break}\) is assigned to the \(B=B_{\rm common}\) crossing
obtained by linear interpolation between the two outputs bracketing the
crossing.

\section{Constraint monitoring and convergence tests}
\label{app:numerics}

\begin{figure}[t]
    \centering

    \begin{minipage}{0.49\textwidth}
        \centering
        \includegraphics[width=\linewidth]
        {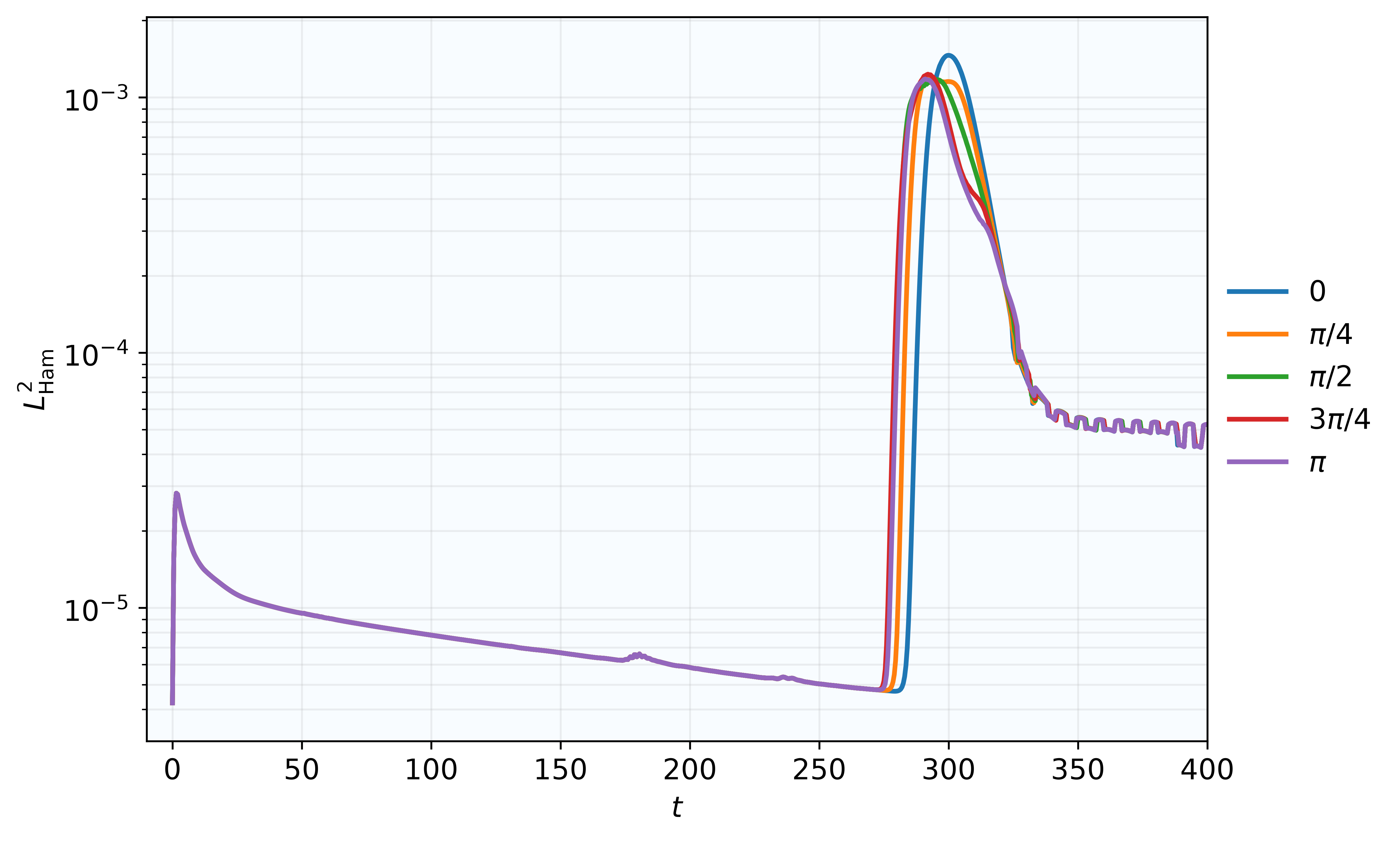}
    \end{minipage}
    \hfill
    \begin{minipage}{0.49\textwidth}
        \centering
        \includegraphics[width=\linewidth]
        {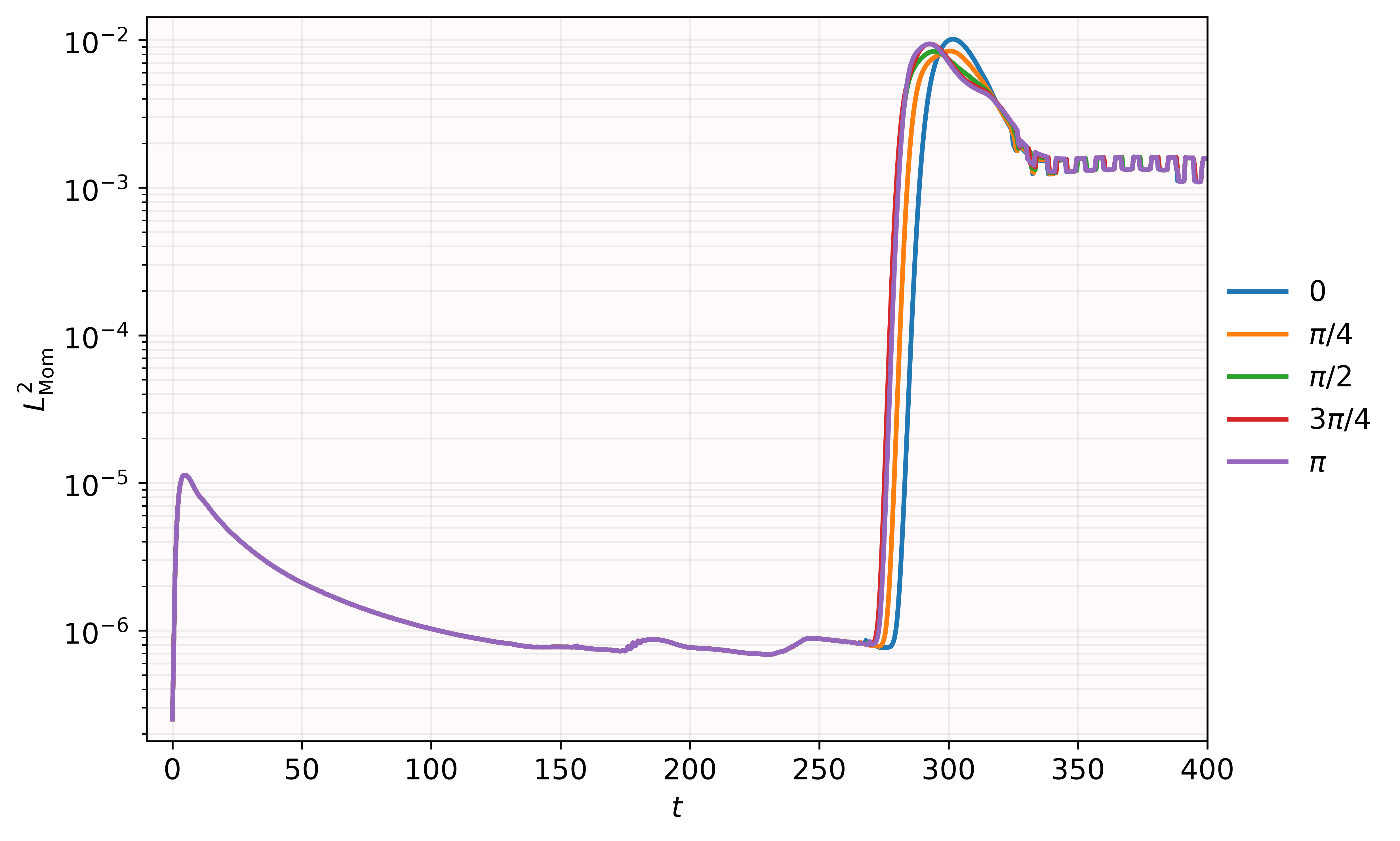}
    \end{minipage}

    \caption{
    Constraint evolution for the benchmark phase scan.
    The left and right panels show the \(L^2\) norms of the Hamiltonian
    and momentum constraints, respectively. The five phase evolutions
    remain at comparable constraint levels throughout the bridge-formation
    and bridge-survival stages. The larger excursion occurs later, during
    the strongly nonlinear evolution.
    }
    \label{fig:benchmark_constraints}
\end{figure}

We monitor the Hamiltonian and momentum constraints throughout the
benchmark phase scan. Figure~\ref{fig:benchmark_constraints} shows their
\(L^2\) norms for the five relative phases. Following the initial
transient, the Hamiltonian and momentum norms remain at roughly
\(10^{-5}\) and \(10^{-6}\), respectively, until the strongly nonlinear
late-time evolution. Their subsequent growth occurs at approximately the
same stage for all five phases, reaching \(O(10^{-3})\) for the
Hamiltonian constraint and \(O(10^{-2})\) for the momentum constraint
before decreasing again.

The bridge-survival interval used in the main analysis,
\(30\leq\tau\leq50\), ends at \(t\simeq260\)--\(266\) for the benchmark
runs and therefore precedes the sharp late-time increase of both
constraint norms.

We repeated the full benchmark phase scan at three base-grid
resolutions,
\[
N=128,\qquad256,\qquad512,
\]
where \(N\) denotes the number of grid points on the coarsest AMR level
covering the full computational domain. The domain length is \(L=512\),
and the same AMR setup is used throughout. The production runs use
\(N=256\).

The diagnostic output cadence changes with resolution. For this
comparison we therefore determine the \(B=0.10\), \(0.35\), and \(0.50\)
crossings directly from the raw \(B(t)\) data by linear interpolation.
For breakup we use the first interpolated downward crossing of
\(B=0.35\) after the strong-bridge level has been reached. These
interpolated times are used only for the resolution comparison; the
event times quoted in the main text retain the sustainment criteria of
Appendix~\ref{app:bridge}. This avoids mixing spatial-resolution effects
with the different physical duration represented by three consecutive
diagnostic outputs.

Table~\ref{tab:resolution_times} lists the characteristic times for all
five phases.

\begin{table}[t]
\caption{
Resolution dependence of the characteristic bridge times for the
benchmark phase scan. A dash indicates that no such downward crossing
occurs over the analysed evolution.
}
\label{tab:resolution_times}
\centering
\begin{ruledtabular}
\begin{tabular}{cccccc}
\(\Delta\theta\) &
\(N\) &
\(t_{0.10}\) &
\(t_{0.35}\) &
\(t_{0.50}\) &
\(t_{\rm break}\) \\
\hline
\(0\)       & 128 & 209.669 & 226.096 & 232.229 & --- \\
            & 256 & 209.686 & 226.100 & 232.211 & --- \\
            & 512 & 209.691 & 226.103 & 232.218 & --- \\[1mm]

\(\pi/4\)   & 128 & 210.330 & 226.778 & 232.731 & --- \\
            & 256 & 210.348 & 226.790 & 232.671 & --- \\
            & 512 & 210.351 & 226.797 & 232.593 & --- \\[1mm]

\(\pi/2\)   & 128 & 211.817 & 227.897 & 236.343 & 268.492 \\
            & 256 & 211.804 & 227.907 & 236.378 & 268.410 \\
            & 512 & 211.665 & 227.893 & 236.382 & 268.400 \\[1mm]

\(3\pi/4\)  & 128 & 214.782 & 231.419 & 238.117 & 263.733 \\
            & 256 & 214.811 & 231.463 & 237.401 & 263.693 \\
            & 512 & 214.819 & 231.462 & 237.401 & 263.678 \\[1mm]

\(\pi\)     & 128 & 215.067 & 231.864 & 238.751 & 263.752 \\
            & 256 & 215.065 & 231.871 & 238.698 & 263.752 \\
            & 512 & 215.065 & 231.874 & 238.565 & 263.751 \\
\end{tabular}
\end{ruledtabular}
\end{table}

The characteristic times are stable across the three resolutions.
The phase ordering of the three upward crossings, as well as the
presence or absence of bridge breakup, is unchanged. Between the
production and highest resolutions, \(N=256\) and \(512\), the largest
shifts over the full phase scan are \(0.14\), \(0.014\), and \(0.13\)
for the \(B=0.10\), \(0.35\), and \(0.50\) crossings, respectively.
For the three phases that undergo bridge breakup, the largest change in
\(t_{\rm break}\) is \(0.015\).

The resolution study therefore leaves the characteristic bridge times,
the ordering of the upward crossings, and the breakup classification
unchanged. The event times are stable under refinement, but their small
shifts do not follow a sufficiently regular pattern to support a
reliable Richardson-order estimate. We therefore use this resolution
study to test the robustness of the event timing and bridge-breakup
classification, rather than to assign a formal convergence order.

\clearpage
\bibliography{Ref}

\end{document}